# FUNKI: Interactive functional footprint-based analysis of omics data

Rosa Hernansaiz-Ballesteros[1], Christian H. Holland[1,2], Aurelien Dugourd[1,2], Julio Saez-Rodriguez[1,*]

[1] Heidelberg University, Faculty of Medicine, and Heidelberg University Hospital, Institute for Computational Biomedicine, Bioquant, Heidelberg, Germany

[2] Heidelberg University, Faculty of Biosciences, Heidelberg, Germany

* Corresponding author: pub.saez@uni-heidelberg.de

ORCIDs:
RHB: 0000-0001-8536-8848
CHH: 0000-0002-3060-5786
JSR: 0000-0002-8552-8976

**Abstract**

**Motivation:** Omics data, such as transcriptomics or phosphoproteomics, are broadly used to get a snap-shot of the molecular status of cells. In particular, changes in omics can be used to estimate the activity of pathways, transcription factors and kinases based on known regulated targets, that we call footprints. Then the molecular paths driving these activities can be estimated using causal reasoning on large signaling networks.

**Results:** We have developed FUNKI, a FUNctional toolKIt for footprint analysis. It provides a user-friendly interface for an easy and fast analysis of several omics data, either from bulk or single-cell experiments. FUNKI also features different options to visualise the results and run post-analyses, and is mirrored as a scripted version in R.

**Availability:** FUNKI is a free and open-source application built on R and Shiny, available in GitHub at https://github.com/saezlab/ShinyFUNKI under GNU v3.0 license and accessible also in https://saezlab.shinyapps.io/funki/

**Contact**: pub.saez@uni-heidelberg.de

**Supplementary information:** We provide data examples within the app, as well as extensive information about the different variables to select, the results, and the different plots in the help page.

## 1. Introduction

Multiple methods are conceived to infer the activities of specific processes or molecules using the abundance of known targets from omic data (see Supplementary Table for a list of them). We call them footprint-based methods (Dugourd and Saez-Rodriguez, 2019), and we have develop such tools for transcription factor from transcripts of target genes (Garcia-Alonso *et al.*, 2019), kinases from phosphorylated sites (Wirbel *et al.*, 2018), and pathways from downstream responsive genes (Schubert *et al.*, 2018). These activities can then be used to contextualize large signaling networks by identifying paths that can explain the changes in activities via reverse causal reasoning (Liu *et al.*, 2019; Dugourd *et al.*, 2021).

FUNKI (FUNctional analysis toolKIt) is an user-friendly interface developed in R (Team, 2020), and designed using Shiny (Chang *et al.*, 2020), to analyze omics data using footprint methods. This application provides an interface for the R implementations (Bioconductor packages) for the aforementioned tools. All methods run on bulk data for human samples, and we have shown that they can also be applied to single-cell transcriptomics (Holland, Tanevski, *et al.*, 2020) and mouse (Holland, Szalai, *et al.*, 2020).

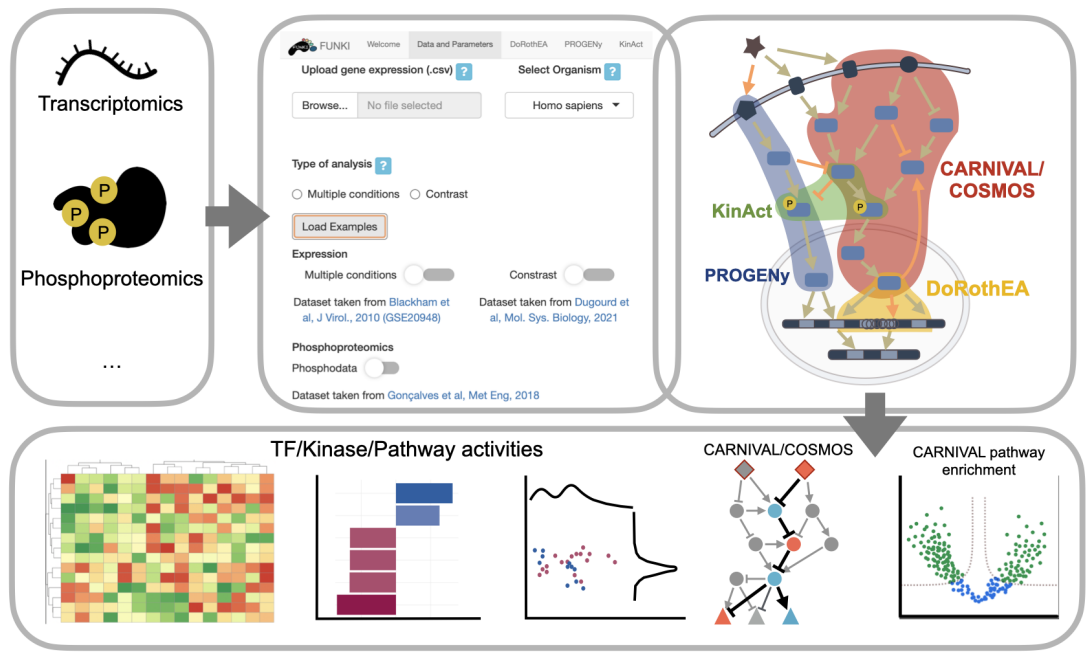

***Figure 1. Graphical overview of analysis and visualisation features provided by FUNKI.***
*FUNKI provides an user interface to upload omics data, and then run DoRothEA, PROGENy, KinAct, CARNIVAL and COSMOS to estimate the activity of pathways, transcription factors, and kinases. The results are visualized in diverse forms.*

## 2. Features

The footprint methods implemented in FUNKI allow users to recover functional insight from several omics data without notions of programming. This application also enhances the analysis with an extended graphic visualisation of the results. Thus, the typical FUNKI pipeline comprises three steps: (i) import user's omic data, (ii) select the analysis accordingly (DoRothEA, PROGENy, CARNIVAL, COSMOS or KInAct), and (iii) visualise the results in tables and graphical representations (Fig. 1).

### 2.1 DoRothEA

DoRothEA (Discriminant Regulon Expression Analysis) is a resource that links transcription factors (TFs) with their downstream targets. TFs activities are computed from gene expression where the regulons (the collection of transcriptional targets for each TF) are the underlying gene-sets.

**2.2 PROGENy**

PROGENy (Pathway RespOnsive GENes) is a footprint method developed to infer pathway activity from gene expression data. The scores are calculated using a linear model with weights based on consensus gene signatures obtained from publicly available perturbation experiments.

**2.3 KinAct**

KinAct is a resource linking kinases to phosphorylation sites. Kinase activity estimation is performed using the same algorithm as DoRothEA to estimate activity scores. Instead of TF-target interactions, KinAct uses collections of kinase-substrate interactions via OmniPath (Türei *et al.*, 2016) and phosphoproteomic data instead of transcriptomic data.

**2.4 CARNIVAL and COSMOS**

CARNIVAL (CAusal Reasoning for Network identification using Integer VALue programming) reconstructs signalling networks from downstream TF activities by finding the upstream regulators. COSMOS is an extension of CARNIVAL that provides a multi-omic network to connect different types of omic data together, including transcriptomics, metabolomics, and phosphoproteomics. Both methods identify coherent mechanistic hypotheses (subnetworks) that explain how the measured deregulation may be reached.

**3. Implementation**

FUNKI is a shiny application developed using R programming language under version 4.0.2 and upgraded to run for 4.1.1 (Team, 2020; Chang *et al.*, 2020). It is directly accessible in the cloud through https://saezlab.shinyapps.io/funki/. The source code is freely available at https://github.com/saezlab/ShinyFUNKI, and it can be run locally in any platform (Windows, macOS and Linux) either downloading the repository or running it directly from GitHub (see https://saezlab.github.io/ShinyFUNKI/ for details).

**4. Conclusion**

FUNKI provides an intuitive user-friendly interface to run footprint methods from different omics. Together with the analysis implementation, FUNKI also incorporates several graphical representations to explore the results from different perspectives. Users with programming skills can take advantage of an extended script-based version of FUNKI for transcriptomic data (https://github.com/saezlab/transcriptutorial).

## 5. Acknowledgements

This project was funded by the European Union's Horizon 2020 research and innovation programme H2020-ICT-2018-2 project iPC – individualized Paediatric Cure (Grant no 826121) and the German Federal Ministry of Education and Research (Bundesministerium für Bildung und Forschung BMBF) MSCoreSys research initiative research core SMART-CARE (031L0212A). Thanks to Nicolas Palacio and the members of Bender's group for testing the beta version and feedback to improve FUNKI.

## 6. Conflict of interests

JSR has received funding from GSK and Sanofi and consultant fees from Travere Therapeutics. RHB has received consultant fees from QuantBio.

## 8. References

Chang,W. *et al.* (2020) shiny: Web Application Framework for R.

Dugourd,A. and Saez-Rodriguez,J. (2019) Footprint-based functional analysis of multiomic data. *Current Opinion in Systems Biology*, **15**, 82–90.

Dugourd,A. *et al.* (2021) Causal integration of multi-omics data with prior knowledge to generate mechanistic hypotheses. *Mol. Syst. Biol.*, **17**, e9730.

Garcia-Alonso,L. *et al.* (2019) Benchmark and integration of resources for the estimation of human transcription factor activities. *Genome Res.*, **29**, 1363–1375.

Holland,C.H., Tanevski,J., *et al.* (2020) Robustness and applicability of transcription factor and pathway analysis tools on single-cell RNA-seq data. *Genome Biol.*, **21**, 36.

Holland,C.H., Szalai,B., *et al.* (2020) Transfer of regulatory knowledge from human to mouse for functional genomics analysis. *Biochim. Biophys. Acta Gene Regul. Mech.*, **1863**, 194431.

Liu,A. *et al.* (2019) From expression footprints to causal pathways: contextualizing large signaling networks with CARNIVAL. *NPJ Syst. Biol. Appl.*, **5**, 40.

Schubert,M. *et al.* (2018) Perturbation-response genes reveal signaling footprints in cancer gene expression. *Nat. Commun.*, **9**, 20.

Team,R.C. (2020) R: A Language and Environment for Statistical Computing.

Türei,D. *et al.* (2016) OmniPath: guidelines and gateway for literature-curated signaling pathway resources. *Nat. Methods*, **13**, 966–967.

Wirbel,J. *et al.* (2018) Phosphoproteomics-Based Profiling of Kinase Activities in Cancer Cells. *Methods Mol. Biol.*, **1711**, 103–132.

**Supplementary table 1**

| Name | Activity type | Available interface | Reference |
|---|---|---|---|
| Viper | TF/Kinase | No | (Alvarez *et al.*, 2016) |
| RoKAI | Kinase | No | (Yılmaz *et al.*, 2021) |
| KARP | Kinase | No | (Wilkes *et al.*, 2017) |
| KSEA | Kinase | No | (Hernandez-Armenta *et al.*, 2017) |
| KSTAR | Kinase | No | (Crowl *et al.*, 2021) |
| INKA | Kinase | No | (Beekhof *et al.*, 2019) |
| KEA3 | Kinase | Yes | (Kuleshov *et al.*, 2021) |
| BART | TF | No | (Wang *et al.*, 2018) |
| TFEA.ChiP | TF | Yes | (Puente-Santamaria *et al.*, 2019) |
| oPOSSUM | TF | Yes | (Kwon *et al.*, 2012) |
| CHEA3 | TF | Yes | (Keenan *et al.*, 2019) |
| MAGICACT | TF | No | (Roopra, 2018) |
| SPEED | Pathway | Yes | (Parikh *et al.*, 2010) |
| SPEED2 | Pathway | Yes | (Rydenfelt *et al.*, 2020) |

*List of other methods to infer the activity of proteins from different omics based on the idea of looking at the target molecules (what we call the footprint-based approach). TF = Transcription Factor.*

## Bibliography

Alvarez,M.J. *et al.* (2016) Functional characterization of somatic mutations in cancer using network-based inference of protein activity. *Nat. Genet.*, **48**, 838–847.

Beekhof,R. *et al.* (2019) INKA, an integrative data analysis pipeline for phosphoproteomic inference of active kinases. *Mol. Syst. Biol.*, **15**, e8981.

Crowl,S. *et al.* (2021) KSTAR: An algorithm to predict patient-specific kinase activities from phosphoproteomic data. *BioRxiv*.

Hernandez-Armenta,C. *et al.* (2017) Benchmarking substrate-based kinase activity inference using phosphoproteomic data. *Bioinformatics*, **33**, 1845–1851.

Keenan,A.B. *et al.* (2019) ChEA3: transcription factor enrichment analysis by orthogonal omics integration. *Nucleic Acids Res.*, **47**, W212–W224.

Kuleshov,M.V. *et al.* (2021) KEA3: improved kinase enrichment analysis via data integration. *Nucleic Acids Res.*, **49**, W304–W316.

Kwon,A.T. *et al.* (2012) oPOSSUM-3: advanced analysis of regulatory motif over-representation across genes or ChIP-Seq datasets. *G3 (Bethesda)*, **2**, 987–1002.

Parikh,J.R. *et al.* (2010) Discovering causal signaling pathways through gene-expression patterns. *Nucleic Acids Res.*, **38**, W109-17.

Puente-Santamaria,L. *et al.* (2019) TFEA.ChIP: a tool kit for transcription factor binding site enrichment analysis capitalizing on ChIP-seq datasets. *Bioinformatics*, **35**, 5339–5340.

Roopra,A. (2018) MAGIC: A tool for predicting transcription factors and cofactors binding sites in gene sets using ENCODE data. *BioRxiv*.

Rydenfelt,M. *et al.* (2020) SPEED2: inferring upstream pathway activity from differential gene expression. *Nucleic Acids Res.*, **48**, W307–W312.

Wang,Z. *et al.* (2018) BART: a transcription factor prediction tool with query gene sets or epigenomic profiles. *Bioinformatics*, **34**, 2867–2869.

Wilkes,E.H. *et al.* (2017) Kinase activity ranking using phosphoproteomics data (KARP) quantifies the contribution of protein kinases to the regulation of cell viability. *Mol. Cell. Proteomics*, **16**, 1694–1704.

Yılmaz,S. *et al.* (2021) Robust inference of kinase activity using functional networks. *Nat. Commun.*, **12**, 1177.